\documentclass[aps,prb,twocolumn]{revtex4} 
\usepackage[]{amsmath,amssymb,latexsym,epsfig}

\newcommand{\coll}{{\it et al. }}
\newcommand{\jncs}{{J. Non-Cryst. Solids }}
\newcommand{\jpcb}{{J. Phys. Chem. B }}
\newcommand{\philb}{{Philos. Mag. B }}
\newcommand{\jachs}{{J. Am. Chem. Soc. }}
\newcommand{\jacrs}{{J. Am. Ceram. Soc. }}
\newcommand{\jpc}{{J. Phys. Chem. }}
\newcommand{\jap}{J. App. Phys. }
\newcommand{\surf}{Surf. Sci. }
\newcommand{\fig}{\figurename }
\newcommand{\ie}{{\it ie }}
\newcommand{\degre}{$^{\rm o}$}

\begin{document}

\title{Sodium diffusion through amorphous silica surfaces: A molecular dynamics 
study}

\author{Micha\"el Rarivomanantsoa$^{a}$, Philippe Jund$^b$ and R\'emi Jullien$^c$}

\affiliation{$^a$Dynamique et Thermodynamique des Milieux Complexes - UMR 5569 Hydrosciences- Universit\'e
Montpellier~2, Place E. Bataillon Case MSE, 34095 Montpellier Cedex 5 France}

\affiliation{$^b$Laboratoire de Physicochimie de la Mati\`ere Condens\'ee - 
Universit\'e Montpellier 2, Place E. Bataillon Case 03, 34095 Montpellier Cedex 5 
France}

\affiliation{$^c$Laboratoire des Verres - Universit\'e Montpellier 2, Place E. 
Bataillon Case 69, 34095 Montpellier Cedex 5 France}
\date{}

\begin{abstract}
We have studied the diffusion inside the silica network of sodium atoms initially 
located outside the surfaces of an amorphous silica film. We have focused our 
attention on structural and dynamical quantities, and we have found that the local 
environment of the sodium atoms is close to the local environment of the sodium 
atoms inside bulk sodo-silicate glasses obtained by quench. This is in agreement 
with recent experimental results. 
\end{abstract}

\maketitle

\narrowtext

\section{Introduction}
Many characteristics of materials such as mechanical resistance, adsorption, 
corrosion or surface diffusion depend on the physico-chemical properties of the 
surface. Thus, the interactions between the surfaces with their physico-chemical 
environment are very important, and in particular for amorphous materials which 
are of great interest for a wide range of industrial and technological applications 
(optical fibers coating, catalysis, chromatography or microelectronics). Therefore 
a great number of studies have for example focused on the interactions between 
the amorphous silica surfaces with water, experimentally \cite{exp1} and by 
molecular dynamics simulations \cite{md1,md12,bakaev,md10,md11,md13}. 

On the other hand, the sodium silicate glasses entail great interest due to their 
presence in most of the commercial glasses and geological magmas. They are also 
often used as simple models for a great number of silicate glasses with more 
complicated composition. The influence of sodium atoms on the amorphous silica 
network is the subject of numerous experimental studies: Raman spectroscopy 
\cite{brawer,millan}, IR \cite{brawer,wong}, XPS \cite{bruckner,sprenger} and NMR 
\cite{sprenger,silver} from which we have informations about neighboring distances,
bond angle distributions or concentration of so-called Q$^{n}$ tetrahedra. In order 
to improve the insight about the sodium silicate glass structure and to obtain a 
good understanding of the role of the modifying Na$^+$ cations, Greaves \coll have 
used new promising investigation techniques like EXAFS and MAS NMR 
\cite{baker,greaves}. Despite all these efforts, the structure of sodo-silicate 
glasses is still a subject of debate. Another means to 
give informations about this structure is provided by simulations, by either {\it ab initio} 
\cite{ispas} or classical \cite{soules,vessal,smith,oviedo,horbach,jund} molecular dynamics (MD). 
In the present work, we are using classical MD simulations, 
but {\it a contrario} to previous simulations, the sodium atoms are not located
before hand inside the amorphous silica sample.

Recent experimental studies of the diffusion of Na atoms initially placed at the 
surface of amorphous silica, using EXAFS spectroscopy \cite{mazzara}, showed that the 
Na atoms diffuse inside the vitreous silica and once inside the amorphous silica 
network, the local environment of the Na atoms is characterized by a Na~-~O distance $d_{{\rm 
Na-O}} = 2.3$ \AA\ and by a Na~-~Si distance $d_{{\rm Na-Si}} = 3.8$ \AA. These 
values are close to the distances characterizing the local environment of Na atoms 
in sodium silicate glasses obtained by quench.

In this study we have used classical MD simulations in order
to reproduce the diffusion of sodium atoms inside a silica matrix and to check  
that the local environment of the sodium atoms is close to what is found for 
quenched sodo-silicate glasses. The sodium atoms have been inserted at the surface 
of thin amorphous silica films under the form of Na$_2$O groups in
order to respect the charge neutrality. 

\section{Computational method}

To simulate the interactions between the different atoms, we use a generalized 
version \cite{kramer} of the so-called BKS potential \cite{bks} where the 
functional form of the potential remains unchanged:
\begin{equation*}
{\cal{\phi}}\left(\left|\vec{r}_j-\vec{r}_i\right|\right) = 
\frac{q_iq_j}{\left|\vec{r}_j-\vec{r}_i\right|}
-A_{ij}\exp\left(-B_{ij}\left|\vec{r}_j-\vec{r}_i\right|\right)
-\frac{C_{ij}}{\left|\vec{r}_j-\vec{r}_i\right|^6}.
\end{equation*} 
The potential parameters $A_{ij}$, $B_{ij}$, $C_{ij}$, $q_i$ and $q_j$ involving the silicon 
and oxygen atoms (describing the interactions inside the amorphous silica network) are extracted 
from van Beest \coll \cite{bks} and remain unchanged (in particular 
the partial charges q$_{{\rm Si}} = 2.4\rm e$ and q$_{{\rm O}} = -1.2$e are not modified).
The new parameters, devoted 
to describe the interactions between the sodium atoms and the silica network are given 
by Kramer \coll \cite{kramer} and are adjusted on {\it ab initio} calculations of zeolithes 
except the partial charge of the sodium atoms whose value q$_{{\rm Na}} = 0.6\rm e$ is chosen in order 
to respect the system electroneutrality. 
However, this sodium partial charge does not reproduce the short-range forces and to this purpose, Horbach 
\coll \cite{horbach} have proposed to vary the charge q$_{{\rm Na}}$ as follows:
\begin{eqnarray*}
q_{{\rm Na}}(r_{ij})&=&
   \left\{
      \begin{array}{ll}
         0.6\left(1+\ln\left[C\left( r_c-r_{ij}\right)^2+1\right]\right) & 
         r_{ij} < r_c\\
         0.6 & r_{ij} \geqslant r_c
      \end{array}
   \right.
\end{eqnarray*}
where $r_{ij}$ is the distance between the particles $i$ and $j$. The parameters 
$C$ and $r_c$ are adjusted to obtain the experimental structure factor of 
Na$_2$Si$_2$O$_5$ (NS2)
%
and their values are included in Ref \cite{horbach}.
%
It is important to note that using this method to model 
the sodium charge, the system electroneutrality is respected for large distances 
(in fact for distances $r \geqslant r_c$). Next we assume that the modified BKS 
potential describes reasonably well the system studied here, for which the sodium 
atoms are initially located outside the amorphous silica sample. In addition other 
simulations have shown that this interatomic potential is convenient for various compositions, 
in particular for NS2, NS3 (Na$_2$Si$_3$O$_7$) \cite{horbach} and NS4 (Na$_2$Si$_4$O$_9$) 
\cite{jund} and we assume it is adapted to model any concentration of modifying 
Na$^+$ cations inside sodo-silicate glasses.

\begin{figure}[h]
\centerline{\epsfxsize280pt{\epsffile{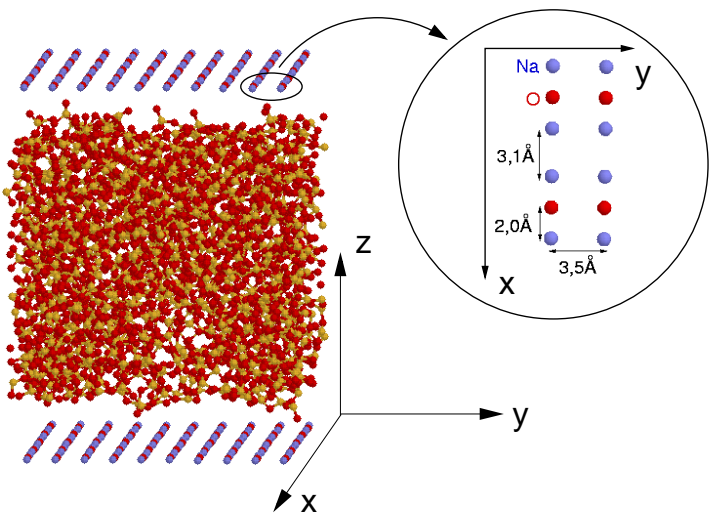}}}
\caption{\label{figure1} Initial configuration: two layers containing 50 
{\rm Na}$_2${\rm O} groups each placed at 4 {\rm \AA} of the
two free surfaces of an amorphous 
silica film. The way the {\rm Na}$_2${\rm O} groups are arranged within the layers 
is shown in the blow up.}
\end{figure}

Our aim here is to obtain a sodo-silicate glass by deposition of sodium atoms at 
the amorphous silica surface, as it was done experimentally \cite{mazzara}. Using 
the {\it modus operandi} described in a previous study \cite{mr} we have generated 
Amorphous Silica Films (ASF), each containing two free surfaces perpendicular 
to the $z$-direction. These samples have been made by breaking the periodic boundary 
conditions along the $z$-direction,
%
normal to the surface, thus creating two free surfaces located at $L/2$ and $-L/2$ with
$L=35.8$ \AA. In order to evaluate the Coulomb interactions, we used a two-dimensional 
technique based on a modified Ewald summation to take into account the loss of periodicity 
in the $z$-direction. For further technical details see Ref \cite{mr}.
%
Then, instead 
of initially positioning the sodium atoms inside the silica matrix, like it was done 
before \cite{soules,huang,smith,jund,oviedo}, we have deposited 50 Na$_2$O groups inside 
two layers located at a distance of 4 \AA\ of each free surface as depicted in 
\fig~\ref{figure1}.

Within the layers, the Na$_2$O groups are assumed to be linear, with $d_{{\rm 
Na-O}}=2$~\AA\ \cite{elliott}, and arranged on a 
pseudoperiodic lattice represented in the zoom of \fig~\ref{figure1}. Hence the 
system is made of 100 Na$_2$O groups for 1000 SiO$_2$ molecules, corresponding to
a sodo-silicate glass of composition NS10 (Na$_2$Si$_{10}$O$_{21}$) and contains 
3300 particles. Since our goal is to study the diffusion of the sodium atoms 
placed at the amorphous silica surfaces, we fixed the initial temperature of the 
whole system at 2000~K. Indeed, the simulations of Smith \coll \cite{smith} and 
Oviedo \coll \cite{oviedo} of sodo-silicate glasses have shown that
there is no appreciable 
sodium diffusion for temperatures below $\approx$ 1500~K. On the other hand, 
it is worth noticing that Sunyer \coll \cite{sunyer} have found a simulated glass 
transition temperature $T_g \simeq 2400$~K for a NS4 glass. Therefore we thermalized 
the sodium layers at 2000~K and placed them at the ASF surfaces, also thermalized at 2000~K.

We have used repulsive 'walls' at $z=-30$ \AA\ and $z=+30$ \AA\ 
in order to avoid that some Na$_2$O groups evaporate along the $z$ direction, 
where the periodic boundary conditions are no more fulfilled. The energy of the 
repulsive 'walls' has an exponential shape, $E=E_0\exp\left [-\left ( z_w-z\right )/\sigma \right ]$ 
where $z_w$ is the wall position, $\sigma=0.1$~\AA\ the distance for which the repulsion energy 
is diminished by a factor $e$ and $E_0=10$~eV the repulsion energy at the plane $z=z_w$. 
The value of 30 \AA\ for $z_w$ was chosen in order to place the repulsive walls 
at a reasonable distance from the Na$_2$O layers and not too far from the ASF 
surfaces.

We have then performed classical MD simulations, with a timestep $\Delta t=0.7$ fs, 
using ten statistically independent samples. Since the interactions between the surfaces and 
the Na$_2$O layers are relatively weak, some Na$_2$O groups may evaporate just 
before being reflected toward the thin films by the repulsive walls. During this 
time frame, the system temperature increases up to a temperature of 
approximately 2800~K. 
%
As described by Athanasopoulos \coll \cite{athanasopoulos} this temperature rise is likely
due to the approach of the adatoms to the surface of the substrate, dropping in the potential 
well of the substrate atoms, thus increasing their kinetic energy. This kinetic energy is then 
transmitted to the substrate with the adsorption. Contrarily to Athanasopoulos \coll 
\cite{athanasopoulos} and Webb \coll \cite{webb} we have not dissipated this energy excess 
with a thermal sink region, in fact we have not controlled the temperature at all.
%
Using this device we are able to perform MD simulations of the diffusion of the sodium atoms 
deposited {\it via} Na$_2$O groups at the ASF surfaces. In the following section, we will 
present the structural and dynamical characteristics of the sodo-silicate film (SSF) obtained
in this way.

\section{Results}

\begin{figure}[h]
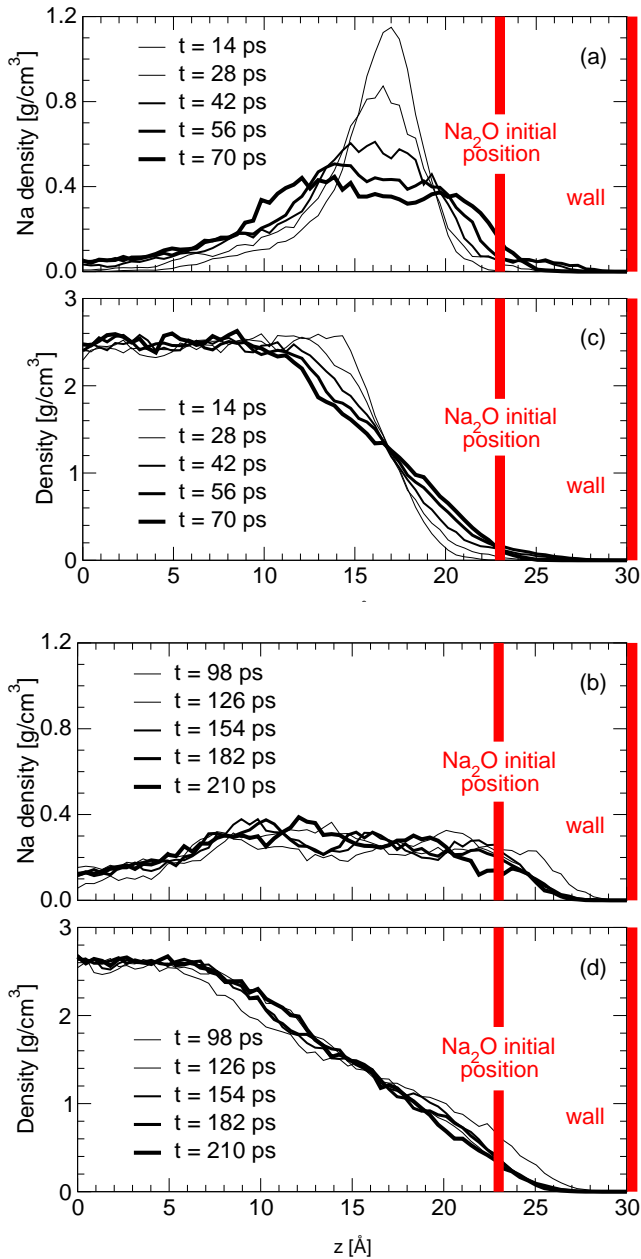

\begin{center}
\begin{minipage}{0.45\textwidth}
\epsfxsize240pt{\epsffile{figure2a.eps}}
\end{minipage}
\begin{minipage}{0.45\textwidth}
\epsfxsize240pt{\epsffile{figure2b.eps}}
\end{minipage}
\end{center}
\caption{\label{figure2} Time evolution of the Sodium (a) and (b) and total (c) and (d) density 
profile.}
\end{figure}

The observation of the sodium diffusion in the silica network is an important goal 
of this molecular dynamics simulation. This can be carried out by analyzing the behavior of the 
density profiles along the normal direction to the surface (the $z$-direction). 
The density profiles represent the mass densities within slices, of thickness 
$\Delta z=0.224$~\AA, parallel to the surfaces \cite{mr}. The time evolution of the 
sodium density profile is represented in \fig~\ref{figure2}(a) and \fig~\ref{figure2}(b) 
and the time evolution of the total density profile in \fig~\ref{figure2}(c) and 
\fig~\ref{figure2}(d) .

The sodium atoms enter inside the silica network in the time range 14~-~70~ps 
(\fig~\ref{figure2}(a)) and during this time interval the sodium profile exhibits a 
diffusion front. After 42~ps, the sodium atoms are observed in the entire system 
illustrating that they have diffused within the whole ASF, as observed experimentally 
\cite{mazzara}.
%
This result contrasts with that of MD simulations of the diffusion 
of platinum (electrically neutral) \cite{athanasopoulos} and potassium 
\cite{garofalini2,zirl1,zirl2} but is similar to 
that of the diffusion of lithium \cite{garofalini2,zirl1,zirl2} 
at the surface of amorphous silica. Moreover, the sodium profiles (\fig~\ref{figure2}(a) 
and (b)) show outward rearrangement at the surface as already pointed out by Zirl \coll 
\cite{zirl} for glassy sodium aluminosilicate surfaces. This agrees with ion scattering 
spectroscopy \cite{kelso} and simulations of sodium silicate glasses \cite{garofalini} 
in which high concentrations of sodium ions are found at the surface.
For larger times, $t\geqslant 98$~ps represented in \fig~\ref{figure2}(b), the sodium 
density seems to stabilize around a mean value of 0.3~g.cm$^{-3}$ which corresponds to 
$\sim 2$ Na atoms per slice ($\Delta z = 0.224$~\AA\ ). 

The surface of the system is identified as being the large linear region in which the total 
density profile decreases. For the short times (\fig~\ref{figure2}(c)) the 
surface is located approximately in the range 15~-~20~\AA\ and for larger times 
($t\geqslant 98$~ps, \fig~\ref{figure2}(d)), the surface lies in the range 
7~-~25~\AA. Hence, the introduction of the sodium atoms in the ASF is likely to increase 
the surface thickness of the system. On the other hand, as observed for the adsorption of 
platinum atoms on the surface of amorphous silica \cite{athanasopoulos}, the surface 
position does not seem to vary with time. We have also calculated the silicon and 
oxygen density profiles, but since they behave like the total density profile, they 
are not represented here. The non bridging oxygen (NBO) density profile is not 
represented in \fig~\ref{figure2} as well since it is close to the Na density 
profile. 

In the time range 98~-~210~ps, the total and sodium density profiles do not 
evolve with time. In particular, in the region $z \lesssim 5$~\AA\ 
(\fig~\ref{figure2}(d)), the total density value remains fluctuating around a mean 
value of 2.6 g.cm$^{-3}$ and the atom composition is approximately of 370 Si, 760 
O and 40 Na which is usually written Na$_{2}$O\,(SiO$_{2}$)$_{18.5}$ or 'NS18.5' 
(it should be noticed that the above mentioned density is significantly larger
than the one expected for a ``real'' NS18.5 glass ($\approx$ 2.3 g.cm$^{-3}$)). 
Therefore it seems reasonable to consider that the system is in a {\it quasi 
permanent} regime after 210~ps and the forthcoming quantities, structural and 
dynamical, are calculated for the following 70~ps. It is worth remembering that all 
the quantities are determined for a system containing 3300 particles and averaged 
over 10 statistically independent samples.

As usual when studying the structural and dynamical characteristics of free 
surfaces, the system is divided into several subsystems: here six slices of 
equal thickness 10~\AA. But in order to increase the statistics, the contributions 
to the physical quantities of the negative and positive slices are averaged. Hence, 
the system is actually subdivided into three parts, named respectively from the center 
to the surface, interior, intermediate and external region.

\begin{figure}[h]
\begin{center}
\begin{minipage}{0.5\textwidth}
\centerline{\epsfxsize240pt{\epsffile{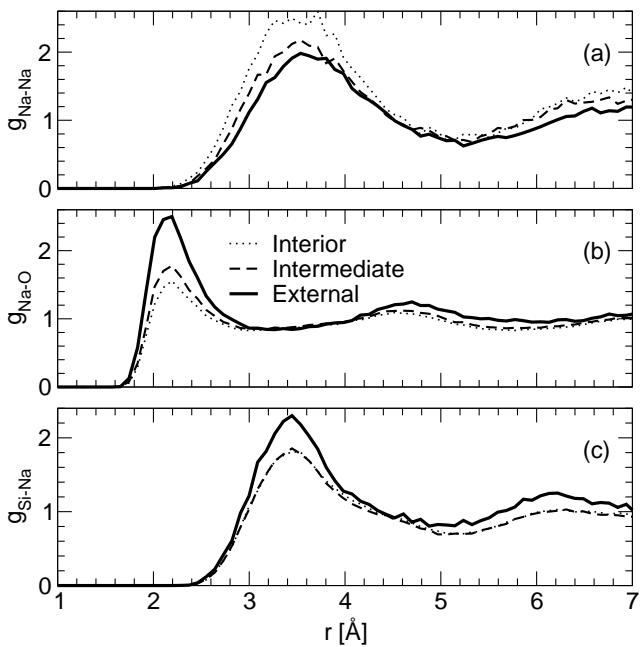}}}
\end{minipage}
\end{center}
\caption{\label{figure3} (a) {\rm Na~-~Na}, (b) {\rm Na~-~O} and (c) {\rm Si~-~Na} 
radial pair distribution functions for the ($\cdots$) interior, (--~--) intermediate 
and (---) external region.}
\end{figure}

In order to improve the characterization of the local environment of the atoms, 
we have calculated the radial pair distribution functions for all the pairs 
$(i,j) \in {\rm [Si,Na,O]}^{2}$ within the three subsystems defined previously. 
The Na~-~Na, Na~-~O and Si~-~Na pair 
distribution functions are represented in \fig~\ref{figure3}(a), \ref{figure3}(b) 
and \ref{figure3}(c) respectively and represent the local environment of
the sodium atoms for $t\geqslant210$~ps. 

At the surfaces of the system the distances are $d_{{\rm Na-O}} \simeq 2.2$ \AA, 
$d_{{\rm Si-Na}} \simeq 3.5$ \AA\ and, due to a lack of statistics, the Na~-~Na 
distance is included in the interval $3.3 \lesssim d_{{\rm Na-Na}} \lesssim 
3.9$~\AA. While slightly smaller, these distances are close to the experimental 
values found by Mazzara \coll ($d_{{\rm Na-O}}=2.3$ \AA\ and $d_{{\rm Si-Na}}=3.8$ 
\AA) \cite{mazzara}. Moreover, the values found in the present work agree with the 
distances, calculated by MD, corresponding to the sodium environment in 
sodo-silicate glasses, of several sodium compositions (NS2 \cite{baker,smith}, 
NS3 \cite{horbach} and 
NS4 \cite{ispas}), obtained by quench. Therefore, as observed experimentally by Mazzara 
\coll \cite{mazzara}, once within the amorphous silica network, the sodium atoms 
have the same local environment as in the sodo-silicate glass obtained by quench. This 
fact is confirmed by the distributions of the $\widehat{{\rm NaONa}}$ and $\widehat{{\rm SiONa}}$ 
bond angles (not shown) which are close to those determined in quenched sodo-silicate 
glasses \cite{oviedo,sunyer}. In particular, the most probable angles are 
90\degre~for $\widehat{{\rm NaONa}}$ and 105\degre~for $\widehat{{\rm SiONa}}$, 
in agreement with the values found by Oviedo \coll \cite{oviedo} and Sunyer \coll 
\cite{sunyer}.\\

The intermolecular distances corresponding to the amorphous silica network structure 
are $d_{{\rm Si-Si}} \simeq$ 3.1 \AA, $d_{{\rm Si-O}} \simeq$ 1.6~\AA\ and $d_{{\rm 
O-O}} \simeq$ 2.6 \AA. On the other hand, the most probable values exhibited by 
the distributions of the $\widehat{{\rm OSiO}}$ and $\widehat{{\rm SiOSi}}$ bond angles 
(not shown) are respectively $\sim 109$\degre~and $\sim 145$\degre. These distances and bond 
angle distributions are very similar to those found experimentally or by MD 
 simulations in bulk amorphous silica. The shoulder exhibited at 2.5 
\AA\ by the distribution $g_{{\rm Si-Si}}$ at amorphous silica surfaces \cite{mr} 
and interpreted as the signature of the twofold rings is not present in the 
SSF. 
%
The absence of the shoulders at 80 and 100\degre~depicted at ASF surfaces
\cite{mr,md1,roder} by 
the $\widehat{{\rm OSiO}}$ and $\widehat{{\rm SiOSi}}$ bond angle distributions 
confirms that the small sized rings have disappeared as suggested by the radial 
pair distributions. This result is expected for R$_2$O (with R belonging to the column I)
adsorption on glassy silicate surfaces and observed {\it via} MD simulations for R = H, 
Li, Na and K \cite{md1,md12,md11,garofalini,garofalini2,zirl1,zirl2}. 
Note that this also occurs for platinum adsorption on sodium aluminosilicate surfaces 
\cite{athanasopoulos,webb} and for H$_2$ adsorption on amorphous silica surfaces \cite{lopez}. 
The small rings like two and threefold rings are known
to be some of the most reactive sites on the surfaces of silicate glasses since they include 
strained siloxane bonds that react with water or other adsorbates \cite{brinker,bunker}.
%

Since the sodium introduction weakens the amorphous silica network, one important 
question is to measure the proportion of non bridging oxygens (NBOs). To this purpose,
oxygen coordination with silicon was calculated. As expected, there is an important
proportion of defective oxygens due to the presence of modifying cations Na$^{+}$
(8.9~\% for the SSF to be compared with 
%
1~\% for the ASF \cite{mr}), illustrated by the similarities between the NBO and Na 
density profiles mentioned previously. Moreover, the NBO concentration is coherent with
the concentration of 11~\% of NBO calculated by MD simulations in a NS9 system 
\cite{huang}. The latter concentration is higher than the one found in this study
because of the greater proportion of Na atoms in NS9 compared to 100 Na$_2$O for 
1000 SiO$_2$ in the present SSF.
%
At the surface ($z \geqslant 20$ \AA), the BOs disappear and correlatively, the 
defective oxygens become preponderant, as observed for 
%
amorphous silica (36.2~\% of NBOs at the SSF surfaces, 15~\% at the ASF surfaces
\cite{mr} and 10~\% at the nanoporous silica surfaces \cite{beckers}). 
%
When analyzing the silicon coordination 
with oxygen, we can state that the silicon atoms remain coordinated in a tetrahedral 
way, revealing that the sodium introduction does not modify the silicon environment. 
The modifications created by the Na$^{+}$ cations are not able to break the 
SiO$_4$ tetrahedra which are very stable in the amorphous silica network. This 
agrees with the usual models for the sodo-silicate glasses, \ie the CRN model of 
Zachariasen \cite{zachariasen} and the MRN model of Greaves \cite{greaves}.

One possible way to analyze more precisely the silicon environment consists in 
calculating the Q$^n$ tetrahedra proportion. A Q$^n$ structure is a SiO$_4$ 
tetrahedra which contains $n$ BOs. The Q$^n$ proportion is often determined 
by NMR experiments \cite{chuang,maekawa}, XPS \cite{sprenger,emerson} or by molecular dynamics 
simulations\cite{smith}, in order to describe the local environment around the silicon atoms. 
In the SSF, the Q$^2$ and Q$^3$ concentrations (6.8~\% and 25.4~\% respectively) 
are  weak compared to those determined for NS2, NS3 and NS4 glasses. This result 
is coherent since the NS2, NS3 and NS4 glasses contain more Na$^+$ cations than 
the SSF studied in this work. At the surface, the Q$^3$ proportion is 45.9~\% 
which is comparable to the proportions obtained by  MD (using the BKS potential) 
in NS2, NS3 and NS4 glasses and to the experimental proportions in NS3 \cite{sprenger,silver} 
and NS4 \cite{maekawa,emerson} glasses. 
Moreover, it is worth noting that some Q$^1$ appear at the surface. In fact, 
these structural entities do not allow to create a network but it is conceivable 
to find those defects forming 'dead ends' at the surface.

A direct method to confirm the previous assumption concerning the disappearance of 
small rings consists in analyzing directly the ring size distribution.
A ring is a
particularly interesting structure because it can be detected using infrared and 
Raman spectroscopy. In particular the highly strained twofold rings result 
in infrared-active stretching modes \cite{bunker,morrow} at 888 and 908 cm$^{-1}$. In 
order to determine the probability $P_n$ for a given Si atom, whose coordinate 
along the normal direction to the surface is in one of the three regions, to be a 
member of a $n$-fold ring we have used the algorithm described in \cite{horbach2}. 
A ring is defined as the shortest path between two oxygen atoms, first neighbors 
of a given silicon atom and made by Si~-~O bonds. The ring size is given by the 
number of silicon atoms contained in the ring.\\

\begin{figure}[h]
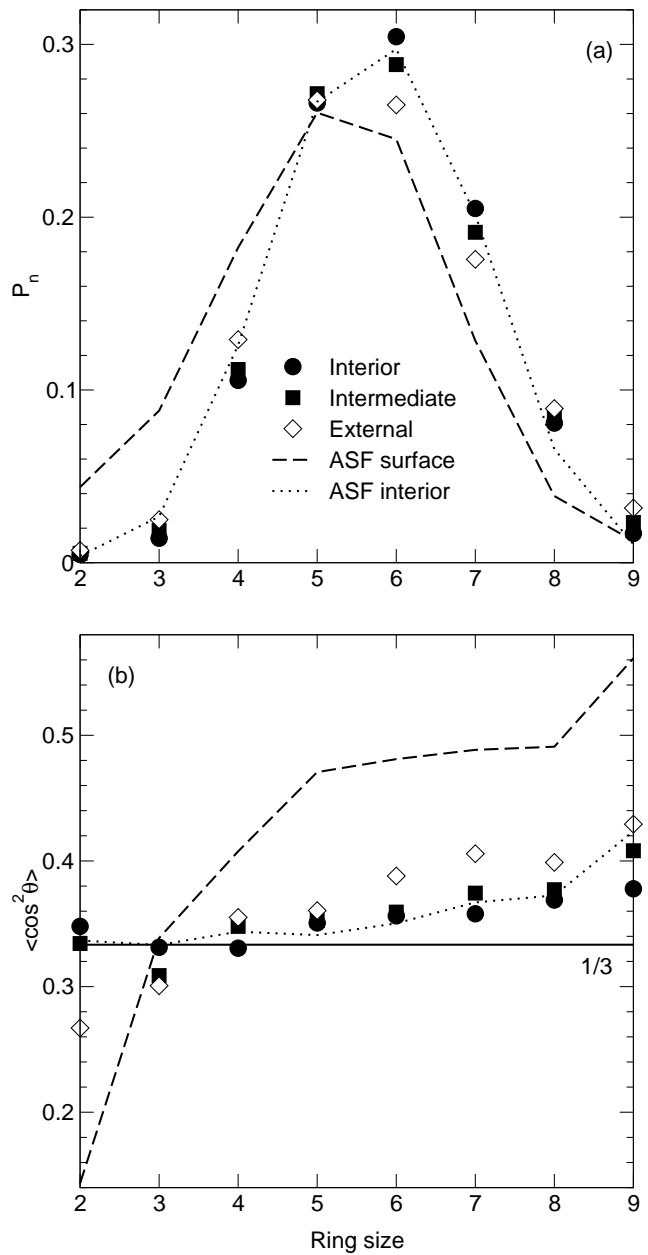

\begin{center}
\begin{minipage}{0.45\textwidth}
\centerline{\epsfxsize240pt{\epsffile{figure4a.eps}}}
\end{minipage}
\begin{minipage}{0.45\textwidth}
\centerline{\epsfxsize240pt{\epsffile{figure4b.eps}}}
\end{minipage}
\end{center}
\caption{\label{figure4} Ring probabilities (a) and mean orientations (b) as a function of their
size for the {\Large$\bullet$} interior, $\blacksquare$ intermediate and 
$\Diamond$ external regions. The same quantities calculated for the ASF surface (--~--) 
and ASF interior ($\cdots$) regions \cite{mr}.}
\end{figure}

The probability $P_n$ is reported in \fig~\ref{figure4}(a) for $n=2,\ldots,9$~ and 
for the three different regions. For comparison, we have also reported $P_n$ 
determined at the ASF surface and interior \cite{mr}. In order to improve the 
medium range order characterization, we have investigated the orientation of the rings 
computing $<\cos^2\theta>$ for a given ring size, within the three regions of the 
system, where $\theta$ is the angle between the normal of the surface and the 
normal of the ring \cite{mr}. The results are reported in \fig~\ref{figure4}(b) for the 
three regions and for $n=2,\ldots,9$ together with the results obtained for the ASF 
surface (dashed line) and interior (dotted line). 

For the three regions, the distributions of the sodo-silicate film (\fig~\ref{figure4}(a)) 
are closer to the ASF interior than to the ASF surface distributions. Particularly, 
the probability of a silicon atom to belong to a small sized ring (2, 3 or 4-fold 
ring) is weak. 
%
This confirms the previous conclusion about the disappearance of the small strained
sized rings which react with the sodium ions during their adsorption at the amorphous
silica surface.
%
As observed for the ring size distributions, the orientation of the rings (\fig~\ref{figure4}(b)), 
in the three regions of the sodo-silicate film is similar to that obtained in the interior 
of the pure silica films \cite{mr} which means that  
%
even at the surface the rings have an isotropic orientation with respect 
to the surface. This is related to the disappearance of the small sized rings at
the surface.
%
Nevertheless, it is worth noting that in the external region the probability of a 
Si atom to belong to a 5-fold ring is greater than the probability to belong to 
a 6-fold ring (\fig~\ref{figure4}(a)) as observed at the ASF {\em surfaces} \cite{mr}. 
In a sense, the 5-fold rings are not affected by the sodium adsorption in contrast
to the small sized rings which disappear with the introduction of the sodium atoms.
%
Also in the external region, the 2 and 3-fold rings 
%
(that are still present) 
%
are oriented perpendicularly to the surfaces (\fig~\ref{figure4}(b)) as observed at 
the ASF surfaces \cite{mr,ceresoli}. 

\begin{figure}[h]
\centerline{\epsfxsize240pt{\epsffile{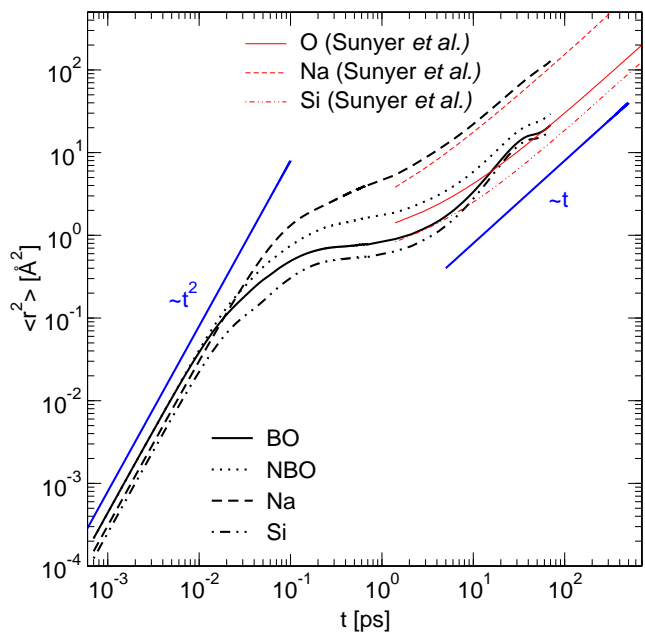}}}
\caption{\label{figure5} Mean square displacements for the {\rm BO} (---), {\rm 
NBO} ($\cdots$), {\rm Na} (--~--) and {\rm Si} ($\cdot$~--~$\cdot$) at 
$\sim$~2800~{\rm K}. The thin lines represent the results obtained by Sunyer \coll 
\cite{sunyer} for a {\rm NS4} system at 3000 {\rm K}.} 
\end{figure}

We have also analyzed the dynamics of the Na$^+$ cations and compared the results  
obtained in the present 'NS10' system with those obtained in a  NS4 glass. To this purpose, 
the mean square displacements (MSD) $\left< r^2(t)\right > = \left<\left|r_i(t)-
r_i(0)\right|^2\right>$ have been calculated for each species composing the SSF. 
\fig~\ref{figure5} represents the MSD for the BO, NBO, Na and Si atoms within the
time frame 0.7~fs~-~70~ps after the first 210~ps together with the MSD calculated by 
Sunyer \coll \cite{sunyer} for NS4 ($t \geqslant 1.5$~ps).

At $\sim$ 2800 K, the MSD of each species exhibits three regimes. For the short 
times, we observe the so-called ballistic regime where $\left< r^2(t)\right > \sim 
t^2$. In this regime, the differences between the species are not really important. For the long 
times, we recognize the so-called diffusive regime in which $\left< r^2(t)\right> 
\sim t$ and in which the Si atoms are the ones that diffuse the less. Similarly 
to the pure ASF \cite{mr}, the NBOs diffuse more than the BOs. The origin of this 
feature lies in the fact that the NBOs form only one covalent bond with the Si 
atoms instead of two for the BOs. The Na atoms diffuse much more (one order of 
magnitude) than the other species. Between the two former regimes, the MSD exhibits 
the so-called $\beta$-relaxation, clearly observable for the oxygen and silicon 
atoms. The phenomenon is not as clear for the sodium atoms, but between $\sim 
10^{-1}$ and $\sim 1$ ps, the Na MSD does not behave like $t$ (sub-diffusion) and it is reasonable 
to consider that the sodium atoms are submitted to the so-called cage effect 
characterizing the $\beta$-relaxation.\\

These results are consistent with the observations made for different sodo-silicate 
systems \cite{smith,oviedo,horbach} for temperatures close to 2800 K. More precisely the 
mean square displacements in the SSF are close to those calculated by Sunyer \coll 
\cite{sunyer} for a quenched NS4 glass at 3000~K. This is somewhat surprising since
in a NS10 system, one expects smaller diffusion constants for all the species compared
to the diffusion constants obtained in a NS4 system. The modus operandi of the present
system (diffusion of the cations through the surface) could explain this observation. 

\section{Conclusion}

This work was motivated by a recent experimental study \cite{mazzara} of the 
local environment of diffusing sodium atoms deposited at the surfaces of thin amorphous silica 
films. We have reproduced numerically this experiment by classical molecular dynamics simulations 
after putting Na$_2$O groups at the surfaces of amorphous silica thin films. 

We have quantitatively analyzed the temporal evolution of the sodium and total densities
and we have checked that the sodium atoms are diffusing inside the amorphous 
silica network. After a given time, the density profiles are no longer evolving, 
and we have calculated the structure of the resulting sodo-silicate glass. 

Our attention has been focused on the local environment of the sodium atoms. Once 
inside the thin film, they are preferentially bound to NBOs as predicted by the 
MRN model of Greaves \cite{greaves}. The distances and bond angle distributions 
show that the sodium atoms have a local environment corresponding to the local 
environment of the sodium atoms in sodo-silicate glasses made by quench, as observed 
by Mazzara \coll \cite{mazzara}. Moreover, the distances $d_{{\rm Na-O}}=2.2$ \AA\ 
and $d_{{\rm Si-Na}}\simeq 3.5$ \AA\ are close to the experimental values.

Concerning the amorphous silica network, we have observed {\it via} the corresponding 
distance and bond angle distributions that its short 
range order is not modified by the introduction of the Na$^+$ cations. We have 
also calculated the ring size distributions and the orientation of the rings
which show that the introduction of the sodium atoms has an influence on the silica 
network but on larger scales compared to those corresponding to the local environment. 
The ring size distributions and the orientations are close to the results obtained in the
bulk of thin amorphous silica films.
%
This is due to the decrease of the proportion of small rings (particularly two and threefold) 
which interact with the Na$_2$O groups since they are known to be highly reactive 
sites for the adsorption of species on amorphous silica surfaces.\\
Finally concerning the dynamics of the different atoms we find results similar to those
obtained in a NS4 glass at a slightly higher temperature.\\
%

{\bf Acknowledgments} Calculations have been performed partly at the ``Centre Informatique 
National de l'Enseignement Sup\'erieur'' in Montpellier.

\end{document}